\documentstyle[11pt]{article}
\newcommand{\blankline}{\vskip .3cm}
\newcommand{\f}{\begin{equation}}
\newcommand{\ff}{\end{equation}}

\begin{document}
\centerline{\LARGE Chern-Simons theory in $11$ dimensions}
\blankline
\rm
\centerline{\LARGE as a non-perturbative phase of {\cal M} theory }
\centerline{Lee Smolin${}^*$}
\blankline
\centerline{\it  Center for Gravitational Physics and Geometry}
\centerline{\it Department of Physics}
 \centerline {\it The Pennsylvania State University}
\centerline{\it University Park, PA, USA 1}
 \vfill
\centerline{Revised, December 30, 1997}
\vfill
\centerline{ABSTRACT}
A Chern-Simons theory in 11 dimensions, which is a piece of the
11 dimensional supergravity action, is considered as a quantum field 
theory in its own right.  We conjecture that it defines a 
non-perturbative phase of M theory in which the metric and
gravitino vanish.  The theory is diffeomorphism invariant
but not topological in that there are local degrees of freedom.
Nevertheless, there are a countable number of momentum variables
associated with relative cobordism classes of embeddings of seven
dimensional manifolds in ten dimensional space.  The canonical
theory is developed in terms of an algebra of gauge invariant 
observables.  We find a sector of the theory corresponding to
a topological compactification in which the geometry of the
compactified directions is  coded in an algebra of functions
on the base manifold.  The diffeomorphism invariant quantum
theory associated to this sector is constructed, and is found
to describe diffeomorphism classes of excitations of three
surfaces wrapping homology classes of the compactified 
dimensions.
 
\blankline
${}^*$ smolin@phys.psu.edu
\eject

\section{Introduction}

String theory has recently evolved in a fascinating direction, leading
to evidence that it represents a class of perturbative expansions
around vacuum states of a non-perturbative theory, whose nature
remains unknown\cite{duality}.    
This conjectured non-perturbative theory has
been called $\cal M$ theory\cite{duality,mtheory}.  
There is evidence that $11$ 
dimensional supergravity plays an important role in its formulation,
at the very least there may be a phase of this theory whose classical
limit corresponds to $11$ dimensional supergravity.  

Despite some provocative suggestions\cite{matrix}, 
the nature of $\cal M$
theory at the fundamental, non-perturbative level remains
unknown.  A non-perturbative theory of quantum gravity must
be one that relies on no background classical metric to give
meaning to either its algebra of observables or perturbation
expansions.  Classical spacetime must emerge from a study of
collective degrees of freedom that describe the critical behavior
of such a theory, it cannot play a role in the formulation of
the theory.   But if geometry thus plays no fundamental 
role, the
theory must be formulated entirely in algebraic and/or
topological terms.
 
In the search for such a theory, two classes of results may offer
useful hints.  The first is topological quantum field theory, which
shows that there are deep relationships between algebra,
representation theory and topology\cite{tqft,4dtqft}.  
In its deepest formulation,
in terms of the theory of tensor categories\cite{categories}, 
$TQFT$ 
reveals new kinds of structures that may play a role in a non-
perturbative 
formulation of a quantum theory of
gravity.  These structures are in fact closely related
to conformal field theory\cite{tqft-cft}.  

Furthermore,  the topological quantum field theories 
are based on finite
dimensional representations of certain observable algebras.
This is very good, as there are independent arguments from the
Bekenstein bound\cite{bekenstein} 
and the holographic hypothesis\cite{thooft,lenny-holo} 
that tell us that
any quantum theory of gravity must have a state space that
decomposes into finite dimensional subspaces corresponding to
measurements made on the boundaries of regions with finite surface 
area\cite{pluralistic}.

The simplest examples of the algebraic structures in $TQFT$ are
spin networks\cite{sn-roger} 
and quantum spin networks\cite{sn-lou}.  It is interesting that
these label the difeomorphism invariant states of quantum
general relativity\cite{sn1} (or of any diffeomorphism invariant 
quantum field theory whose configuration space is based
on a space of connections\cite{sn-baez}.)  
Quantum general relativity is now
understood at the kinematical level (corresponding to spatially
diffeomorphism invariant states\cite{lp1}) 
where it has been found
that the discrete and combinatorial nature of the spin networks 
correspond to the discreteness of quantum geometry at the
non-perturbative 
level\cite{ls-review,volume1}\footnote{For
a longer version of this argument see
\cite{future}  The results of \cite{lp1,sn1,volume1,ham1} have
also been reformulated  in the language of rigorous
mathematical quantum field theory\cite{rigorous,qsdi,qsdii}.}.

At the dynamical level,  despite some non-trivial 
results\cite{ham1,roumen-ham,RB,CRR,qsdi,qsdii},
it is not at all clear that quantum general relativity may have
good critical behavior such as to allow the existence of a good
continuum limit\cite{instability}.  
(That is the theory may have a status
corresponding to random surface theory away from a critical
point: it is well defined microscopically but has no interesting
macroscopic behavior which may be described in terms of
massless fields on a classical background.)  In any case the search
for such critical behavior need not rely on a quantization of
the dynamics of classical general relativity\cite{fmls}.  

Despite this, the results and methods discovered in the study
of non-perturbative quantum general relativity may provide hints
for the construction of a completely non-perturbative formulation
of $\cal M$ theory.  If one takes this point of view there are
a number of possible starting points.  One is to extend the
spin network/TQFT picture to representations of algebras that play
a role in string theory.  Some results in this direction will be reported 
elsewhere, here I would like to describe some results from a
more modest approach, which is to apply the methods of
diffeomorphism invariant quantum field theory directly
to supergravity in $11$ dimensions\footnote{A new formulation of
$11$ dimensional supergravity in terms of new variables analogous
to the Ashtekar variables has been given in \cite{hermannstephan}.}.  

In fact, what is studied here is only a part of that problem.
If one sets the metric and gravitino field to zero, 11 dimensional
supergravity\cite{11dsugra} reduces to a Chern-Simons like theory 
based on
a three form $A_{abc}$.
The action of the theory is simply
\f
S=\int_{{\cal M}_{11}} A \wedge F \wedge F
\label{action}
\ff
where $F=dA$ is a four form and ${\cal M}_{11}$ is an 
eleven dimensional manifold. 

There are several reasons why it is useful to consider the
quantization of this theory before taking on the full eleven
dimensional supergravity.  Even if $11$ dimensional
supergravity corresponds to no good quantum theory, 
this theory may
be of interest, as it may yield a completely non-perturbative
description of the extended objects such as $D$ branes that
play a crucial role in string theory\cite{dbranes}. 
For example, many of the
results concerned with the entropies of extremal and 
near-extremal black holes come down to counting topological
embeddings and intersections of $D$ branes in the compactified
manifolds\cite{strings-bh}.  
It seems likely that there must be a non-perturbative
analogue of this in which these countings reduce to the topology
and combinatorics of diffeomorphism invariant states.  If so,
it is likely that these have to do with the non-perturbative
excitations of the $A_{abc}$ field.  A study of these in the absence
of the metric and gravitino may then yield insights into 
$\cal M$ theory.   

In any case, it is unlikely that anything interesting can come
from the quantization of the metric parts of $11$ dimensional
supergravity unless it has a more interesting formulation analogous
to the JSS\cite{JSS} and CDJ\cite{CDJ} actions of the four dimensional 
theory.  Although there has recently been progress towards that
goal \cite{hermannstephan} by Melosch and Nicolai, that formulation is
still somewhat complicated.  It is then interesting to ask whether
structures associated with the pure $A_{abc}$ sector might give
hints for how to reformulate $11$ dimensional supergravity in
a manner which could lead to more progress with the quantization.

But perhaps the best reason for considering 
the theory (\ref{action}) is that
it may define a phase of $\cal M$ theory.  In non-perturbative
quantum general relativity 
we have learned that in those forms of the
theory amenable to non-perturbative treatment the classical
phase space is extended to include solutions in which the metric
is degenerate, or vanishes altogether\cite{ls-review}.    This  seems
to be the consequence of seeking to describe the theory
directly in terms of a algebras of fields corresponding to the
full geometry, and not just to waves moving on classical
backgrounds.
Although the standard
form of the $11$ dimensional supergravity 
action is not polynomial
in the variables, one can investigate whether the action and
equations of motion have a good limit in which one scales 
the one form
frame fields $e^I_a$ and the gravitino field $\Psi_a$ 
as $e^I_a =t e^I_a$ and
$\Psi_a = t \Psi_a$ and then takes the limit $t \rightarrow 0$.
The theory does have such a limit, which is given by
(\ref{action}).  This suggests that in a non-perturbative treatment
there should be a sector of the state space 
in which $e^I_a$ and $\Psi_a$ vanish\footnote{Of course the limit
breaks supersymmetry.  But $\cal M$ theory must have some
phases that break supersymmetry, otherwise it can have nothing
to do with nature.}.
But it may be expected that  
any non-perturbative theory corresponding to the supergravity
action in $11$ dimensions describes a phase of $\cal M$
theory.  Thus, it is plausible that what is described in this
paper is the a non-perturbative description of a phase
of $\cal M$ theory.

In fact, we find here that the theory (\ref{action}) has lots
of non-trivial structure.  Most significantly, there is a sector
of solutions in which there is a natural compactification in
which the physics of the compactified dimensions is
completely topological.  As a result, the dynamical content
of the compactified directions is entirely 
represented in terms of an algebra of fields on
the uncompactified dimensions.  
This provides a clue that the wealth of
phenomena associated with different compactifications may
eventually be understood in a different way, which is
completely algebraic and combinatoric. 

We may note that the theory (\ref{action}) has certain
similarities to five dimensional Chern-Simons theory.
That has been studied by \cite{FP,focketal,BGH} and some of 
the results can be
extended directly to the present case. Another study of 
higher dimensional Chern-Simons theories, which contains some
results about the eleven dimensional case which are
complementary to those described here is presented in
\cite{maxmark}.

We first sketch the canonical formulation of the theory that
follows from (\ref{action}). In section 4 we introduce
the observable algebra which plays the role of the loop
algebra in quantum gravity and $3$ dimensional Chern-Simons
theory.  In sections 5 and 6 we then restrict the theory to two 
sectors of its solution space where we can find the full observables
algebra and carry out the quantization, which we do in section 7.

\section{The classical theory}

The equations of motion coming from the action (\ref{action}) are,
\f
F \wedge F =0
\label{eom}
\ff

We may note that one class of solutions may be constructed as
follows.  Let us consider an $11$ dimensional manifold
${\cal M}_{11}$ which is locally a product of a $d$ dimensional 
manifolds ${\cal Z}$ and an $11-d$ dimensional manifold
${\cal R}$.  These are coordinatized respectively by
$z^i , i=1,...,d$ and $y^\alpha$, for $\alpha = 1,...,11-d$.  Then
on ${\cal M}$ we may choose  coordinates $x^{\hat{a}}$, with 
$\hat{a} = 1,...,11$ which split as $x^{\hat{a}} = (z^i, y^\alpha )$.
Let us consider a class of three forms $A_{abc}$ such that
\f
F_{\alpha \hat{b}\hat{c} \hat{d}} =0
\label{stflat}
\ff
Then the Bianchi identity $dF=0$ implies that 
$\partial_\alpha F_{ijkl} =0$.
The space of such $F_{\hat{a} \hat{b}\hat{c}\hat{d}}$'s is 
parameterized by a closed form $F_{ijkl}$ on the $d$ dimensional
manifold ${\cal Z}$.  Clearly to be nontrivial $d \geq 4$.
But if $d < 8$ we have a solution to (\ref{eom}).  These are  
not\footnote{Other classes of solutions are considered in
\cite{maxmark}.} the most general solutions to (\ref{eom}).
Other classes of solutions are considered in
\cite{maxmark}.  In this paper we will be primarily interested in
solutions of the form (\ref{stflat}) 
as they are associated to splittings  of ${\cal M}_{11}$ into
products of lower dimensional
manifolds.  In the maximal case we have 
$11=7+4$;  this may be relevant for non-perturbative 
``compactifications"
to four spacetime dimensions.

The theory has two kinds of gauge invariance, eleven dimensional
diffeomorphism invariance, given locally by
\f
\delta_v A = {\cal L}_v A
\ff
and an abelian gauge invariance,
\f
\delta_\lambda A = d \lambda
\ff
where $\lambda$ is a two form.  However, $\lambda$'s of the
form, $\lambda = d\rho$, where $\rho$ is a one form do not 
contribute
to the gauge transformations.  There are then $11 * 10 /2 - 11=44$
degrees of freedom of gauge transformations.  The counting of the
degrees of freedom is subtle and requires the canonical analysis;
the complete counting is carried out in \cite{maxmark} where it is
found that in the general case there are $19$ local degrees of
freedom.  Here we will study some reduced sectors, in which there
are local degrees of freedom which are expressed as functions on
lower dimensional manifolds.

\section{The canonical theory}

We now assume that ${\cal M}_{11} = R \times {\cal N}_{10}$ where
${\cal N}_{10}$ is a compact ten dimensional manifold.  Locally
we may split the coordinates, so that $x^{\hat{a}}= (x^0, x^a)$, where
$a=1,...10$.  From now on all objects are ten dimensional.  
The action
decomposes as,
\f
S=\int dx^0 \int_{{\cal N}_{10}}\left (  A^0 \wedge F \wedge F -
\dot{A} \wedge A \wedge F
\right )
\ff
Here $A$ and $F$ are the pull backs of the corresponding forms to
the spatial sections and $A^{0}$ is a two form, whose components
in local coordinates are $A^0_{bc} = A_{0bc}$.  The
canonical momenta are,
\f
\pi^{abc} = (A \wedge F)^{*abc}
\ff
This gives rise to a set of primary constraints, which are
\f
C^{abc} = \pi^{abc} - (A \wedge F)^{*abc} =0
\ff
The momenta conjugate to $A^0$ vanish as well, which gives rise
to a set of secondary constraints,
\f
H^{ab}= (F \wedge F)^{*ab} =0
\label{H}
\ff
The action is then of the form,
\f
S=\int dx^0 \int_{{\cal N}_{10}}\left ( \pi^{abc} \dot{A}_{abc} -  
A^0_{ab} H^{ab}  \right )
\ff
We first may separate out a set of first class constraints that
generate the abelian gauge transformations. These
are,
\f
G^{bc} = \partial_a C^{abc} \approx \partial_a \pi^{abc} \approx 0
\ff
where $\approx$ means the equality holds on the constraint surface.
 It would be interesting to carry out a full analysis of the 
constraint algebra, but this has not yet been done.
Because of this I will focus on a sector of the theory below.

On the constraint surface $C^{abc}=0$ we may write the symplectic
form 
\f
\omega(\delta_1 A , \delta_2 \pi ) = \int_{\cal N}\delta_1 A_{abc}
\delta_2 \pi^{abc}
\ff
as,
\f
\omega(\delta_1 A , \delta_2 A ) = \int_{\cal N}\delta_1 A 
\wedge \delta_2 A \wedge F
\label{symform}
\ff
$\omega$ may be inverted for generic $F_{abcd}$, not subject
to the constraint $F\wedge F=0$.  To see this, one may view
$F^{* abcdef}$ as a metric on the space of three forms.
Let indices $A,B,C$ represent the $120$ 
three form indices $abc$. Then
$F^{* AB}$ is a symmetric  metric 
that may be inverted generically to find $\rho_{AB}(A)$
such that 
$F^{* AB} \rho_{BC}(A) = \rho_{CB}(A)F^{* BA} =\delta^A_C= 
\delta^{abc}_{def}$.
Then the kinematical Poisson brackets are
\f
\{ A(x)_{abc} , A(y)_{def} \} = \rho_{abcdef}(x) \delta^{10}(x,y)
\ff
Unfortunately, the matrix $F^{ * AB}$ is degenerate on the
constraint surface $F\wedge F=0$ and does not yield the
physical Poisson brackets of the  theory, except in special cases.
The problem of inverting the symplectic form (\ref{symform})
in the presence of the constraint $F \wedge F=0$ is, 
as far as I know, not solved in general; it is related to the
problem of making a complete analysis of the constraints.  
Below I will
discuss how this may be done in one sector of the theory.

\section{Surface variables and algebra}

To quantize the theory, we may try to follow the method that 
worked
in lower dimensional Chern-Simons theory, gauge theories and 
general
relativity and construct an algebra of gauge invariant variables
associated with embeddings of submanifolds in ${\cal N}_{10}$ of
various dimensions\footnote{The quantization of antisymmetric
tensor gauge fields in terms of surface observables was
considered previously in \cite{rodolfo-surface,ls-surface}.}.  
To begin with we construct a set of
variables associated with three dimensional surfaces 
embedded
in ${\cal N}_{10}$.  For every
such surface $\gamma$, we may define an observable
\f
T[\gamma ] \equiv e^{ \int_\gamma A} .
\ff
Conjugate to this we have a momentum variable, associated with
compact seven dimensional submanifolds $S$.  This is
\f
\pi [S] \equiv \int_S \pi^*
\ff
The Poisson bracket between them involves the oriented intersection
number $I[\gamma , S]$ between the three and seven dimensional 
submanifolds.
\f
\{ T[\gamma ] , \pi [S] \} = I[\gamma ,S] T[\gamma ]
\ff
It is easy to see that on the constraint surface most of the momenta
are not independent.  Instead, let $S^\prime $ be cobordic to $S$ 
relative
to ${\cal N}$.  This means that there is an eight dimensional
submanifold $R$ of ${\cal N}_{10}$ such that 
$\partial R = S \cup \tilde{S}^\prime$ (where $\tilde{S}$ is $S$ with 
the
reversed orientation.)

Then we have,
\f
\pi [S] - \pi [S^\prime ] \approx \int_S A \wedge F - 
\int_{S^\prime} A \wedge F
= \int_R F \wedge F \approx 0
\label{cobord}
\ff

It follows that there is an independent momentum variable
$\pi[S]$ associated to each relative cobordism class of seven 
manifolds
$S$ in ${\cal N}$.

At the same time, the theory has local degrees of freedom, as
we may exhibit sets of solutions labeled by continuous parameters,
as we described above. 
We now turn to a study of a sector of the theory in which
we can see how the interplay of a finite number of momentum
variables with continuous spaces of solutions works out.

\section{Topological sectors}

We now consider several sector of solutions on which we will be able
to construct the Poisson brackets.  In this section we
will discuss a sector of solutions that is topological, in the
sense that there are a finite number of physical degrees of
freedom.  In the next section we will see that we can also have
sectors with local degrees of freedom.  

We will assume that locally
${\cal N}_{10}= {\cal Z}_{d} \times {\cal Y}_{10-d}$ for compact
manifolds ${\cal Z}_{d}$ and  ${\cal Y}_{10-d}$ of the indicated
dimensions.
We assume we have local coordinates 
$z^i, i=1,...,d$ on ${\cal Z}_{d}$  and $y^\alpha , \alpha = 1,...,10-d$ 
on  ${\cal Y}_{10-d}$.   Thus, locally the
coordinates $x^a$ on ${\cal N}_{10}$ can be split
as $x^a =(z^i , y^\alpha )$.  Globally, we will require that
${\cal N}_{10}$ is a bundle over ${\cal Z}_{d}$ fibered
by $ {\cal Y}_{10-d}$, with projection map $\pi$.
We will restrict attention to a sector of solutions
to the constraint (\ref{H}) of the form,
\f
F_{\alpha abc}=0
\label{flatness}
\ff
The Bianchi identities $dF=0 $ then imply
$\partial_\alpha F_{ijkl}=0$.  This sector of the  
solution space is then labeled by closed $4$-forms 
on ${\cal Z}_d$ together with a set of
functions $\phi[\tilde{\gamma}](z)$ on ${\cal Z}_d$, which
are defined as follows.
Each function is labeled by the $\tilde{\gamma}$, which are
the   homology
classes of a three manifold in
${\cal Y}_{10-d}$.     Given a compact
three manifold $\gamma^\alpha (\sigma) $ (with coordinates
$\sigma^I , I=1,2,3$)
in ${\cal Y}_{10-d}$ we have a $d$ parameter set of manifolds
in ${\cal N}_{10}$, 
$\gamma^a (\sigma ) (z) = (\gamma^\alpha (\sigma) , z^i )$.
Each of these are three surfaces embedded in the fiber over
the point $z \in {\cal Z}$.

We then may define the functions
\f
\phi[\tilde{\gamma}](z) \equiv T[\gamma (z)]
\ff
By (\ref{flatness}) these are function only of the
homology class of its embedding in ${\cal Y}_{10-d}$.
Thus, on each 
${\cal Y}_{10-d}$ fiber the degrees of freedom are topological.  

In some cases we can find the Poisson brackets of these functions.
To do this let us consider the behavior of the symplectic form
(\ref{symform}) on this sector of solutions.    $\omega$ is degenerate
and block diagonal, the only non-zero entries are 
\f
\omega(\delta_1 A_{\alpha \beta \gamma } ,
\delta_2 A_{\delta \epsilon \phi}) =
\int \delta_1 A_{\alpha \beta \gamma } 
\delta_2 A_{\delta \epsilon \phi} 
F^{*\alpha \beta \gamma \delta \epsilon \phi}
\ff
where 
$F^{*\alpha \beta \gamma \delta \epsilon \phi} =
\epsilon^{\alpha \beta \gamma \delta \epsilon \phi ijkl}F_{ijkl}$.
We will now concentrate on the simplest case, which is $d=4$.   
We then have one non-trivial component of $F_{ijkl}$, which is
\f
F_{ijkl} = \epsilon_{ijkl}\tilde{\Psi} (z) \ \ \ with \ \ \ \
\tilde{\Psi}(z) = {1 \over 24} \epsilon^{mnop}\partial_m A_{nop}
\ff
We note that $\tilde{\Psi}$ is a density on ${\cal Z}$, which
may be set to a constant by a four dimensional diffeomorphism.
Thus there are no local degrees of freedom from the $A_{ijk}$.
This can also be seen from counting, there are four $A_{ijk}$ but
these are eliminated by local gauge transformations and 
four dimensional diffeomorphisms (which are not independent.).

There are global degrees of freedom associated to the $A_{ijk}$,
one associated to each of the third homology classes of ${\cal Z}$.
However, these have vanishing Poisson brackets with the other
observables, and so just label superselection sectors of the 
theory\footnote{Note that this is because we have restricted
the theory to the sector defined by (\ref{flatness}) and $d=4$.}

From now on we will assume that $\tilde{\Psi} \neq 0$. Thus we
are working only in the sector of the phase space defined by
(\ref{flatness}) and the nonvanishing of $\tilde{\Psi}$.  
As there are no other non-vanishing components of the symplectic
form, $\omega$, we may invert it on its non-degenerate subspace,
to find the Poisson brackets.  The only non-vanishing components
are,
\f
\{ A_{\alpha \beta \gamma}(y,z) , A_{\delta \epsilon \phi }
(y^\prime , z^\prime ) \} = {1 \over \tilde{\Psi} (z) } 
\epsilon_{\alpha \beta \gamma \delta \epsilon \phi}
\delta^4 (z,z^\prime ) \delta^6 (y , y^\prime )
\label{toppb}
\ff
We may then integrate to find the Poisson brackets among
the surface variables.  To do this we may use the product structure
to get an embedding of a three surface
$\hat{\gamma}^i (\sigma )$ in
${\cal Y}$ from every three surface 
embedding in $\gamma^a (\sigma) $ in ${\cal N}$.    Similarly,
we have an embedding $\hat{\hat{\gamma}}$ in the base
$\cal Z$.  
Then we have,
\f
\{ T[\gamma ], T[\gamma^\prime ] \} = \sum_{\sigma^*_i}
T[\gamma ] T[\gamma^\prime ] 
Int_{\cal Y}[\hat{\gamma} , \hat{\gamma}^\prime ]
{1 \over \tilde{\Psi}(\hat{\hat{\gamma}}(\sigma^*_i ))}
\delta^4 ((\hat{\hat{\gamma}}(\sigma^*_i ),
(\hat{\hat{\gamma}}^\prime (\sigma^{\prime*}_i ))
\ff
where $\sigma^*_i$ are the coordinates of points of
intersection of the two surfaces and $i$ labels the intersections
when there is more than one.  Here
$Int_{\cal Y}[\hat{\gamma} , \hat{\gamma}^\prime ]$
is the intersection number of two three surfaces in the 
six dimensional space $\cal Y$, given by
\f
Int_{\cal Y}[\gamma ,  \gamma^\prime ]
= \int_\gamma d^3\gamma^{\alpha \beta \gamma}(\sigma )
 \int_\gamma d^3\gamma^{\prime \delta \epsilon \phi}
(\sigma^\prime )
\epsilon_{\alpha \beta \gamma \delta \epsilon \phi}
\delta^6(\gamma (\sigma), \gamma^\prime (\sigma^\prime ))
\ff

Let us now take the loops to lie entirely in the fibers.  In this case
we have a four parameter family of loops $\gamma^i (z)$ for
each loop $\gamma^i$ in $\cal Y$.  We then have Poisson brackets
\f
\{ \phi[\tilde{\gamma}](z) , \phi[\tilde{\gamma}^\prime ](z^\prime ) 
\}
= Int_{\cal Y} [ \tilde{\gamma},  \tilde{\gamma}^\prime ]
{1 \over \Psi(z)}\delta^4 (z,z^\prime )
\phi[\tilde{\gamma}](z) \phi[\tilde{\gamma}^\prime ](z^\prime )
\ff
Here $Int_{\cal Y} [ \tilde{\gamma},  \tilde{\gamma}^\prime ]$
is a the oriented intersection number of the homology classes
of embeddings of three surfaces in $\cal Y$.  

We may now make use of the result that there is a basis
for $H^3 ({\cal Y}_6 )$, consisting of conjugate pairs,
$(\gamma_I, \pi^J )$, of homology classes of three surfaces 
such that 
$Int_{\cal Y}[\gamma_I , \pi^J]= \delta_I^J$.  It is then convenient
to choose a corresponding set of canonical fields
\f
\phi_I(z) = \phi[\tilde{\gamma}_I](z)  
\label{phiz}
\ff
and momenta (which are densities in $\cal Z$),
\f
\tilde{\pi}^J(z) = \tilde{\Psi} (z) \int_{{\pi}^J (z)}A 
\label{piz}
\ff
These satisfy the canonical commutation relations,
\f
\{  \phi_I (z) , \tilde{\pi}^J (z^\prime ) \} = \delta_I^J 
\delta^4  (z, z^\prime )\phi_I (z)
\label{algebra}
\ff

However, now we must recall the condition (\ref{flatness})
which tells us that the observables 
$\phi[\tilde{\gamma} ](z)$ do not actually depend on $z$.
The reason is that because of (\ref{flatness}) and the 
product structure of the manifold,
\f
\int_{\tilde{\gamma} (z)}A =\int_{\tilde{\gamma} (z^\prime)}A .
\ff
Thus, this sector of solutions is actually a topological field theory
with a finite number of degrees of freedom.  To exhibit them
explicitly, we may write the conjugate momenta as
\f
\Pi^I = \int_{{\cal Z}_4} \tilde{\pi}^I
\label{bigpi}
\ff
so that there is one momentum variable for each conjugate pair
of homology classes $H^3({\cal Y}_6)$.

The variables these are conjugate to may just be taken to be
\f
\phi_I = \phi_I (z)
\ff
for any $z$, as they are all equal.
The observables algebra is then,
\f
\{  \phi_I  , {\Pi}^J  \} = \delta_I^J 
\phi_I 
\label{algebra2}
\ff
Apart from this  single commuting variable $\int_{{\cal Z}_4} F$
this is the complete observable algebra of the degrees of freedom
of the sector defined by (\ref{flatness}) with $d=4$ and
$\tilde{\Phi} \neq 0$.

\section{Quasi-topological sectors}

By counting the full eleven dimensional Chern-Simons 
theory has local degrees of freedom, but these
are not seen in the sector we have just considered. This means
that the condition (\ref{flatness}) defines too small of a set
of solutions to the contraints (\ref{H}).  
To free up the local degrees of freedom of the theory we must
study a less restrictive set of solutions.  It is not easy to study
the general case, because of the difficulty of inverting the 
full symplectic form (\ref{symform}) in the presence of (\ref{H}).
But it is not hard to consider a somewhat less restrictive class
of solutions, which has local degrees of freedom. 

We will keep the same topological conditions we used in the last
section.  The base space 
${\cal Z}$ is then four dimensional and the fibers
${\cal Y}$ are six dimensional.   But to define the
flatness condition we first split the six
coordinates $y^\alpha$ into two sets of three
\f
y^\alpha = (y^A , y^{\bar{A}} )
\ff
with $A,\bar{A}=1,2,3$.  We will assume that the splitting can
be made locally so that there are three surfaces in the ``coordinate"
homology classes $\gamma_I$ which are always in the 
constant $y^A$  surfaces, while there are 
representatives of the  ``conjugate" homology classes
$\pi^J$ which are always in the constant $y^{\bar{A}}$ surfaces.
For example, we may take ${\cal Y}=S^3 \times S^3$, with
$\gamma$ and $\pi$ wrapping respectively the first and second
$S^3$ in which case $y^{\bar{A}}$ are coordinates of the first
$S^3$ and $y^A$ are coordinates of the second one.  

We begin looking for a more general class of solutions by eliminating
some degrees of freedom by gauge fixing.  First, since there are no
local degrees of freedom in the $A_{ijk}$ we will fix them to
a constant value $A_{ijk}^0$.  This can be done by a combination
of gauge transformations involving $\lambda_{ij}(z)$ and
diffeomorphisms of $\cal Z$.  Thus, $\delta A_{ijk}=0$.
Then we may fix many of the mixed components of $A_{abc}$
to vanish by a gauge transformation.  For example, if we define
$A^A_{i}=\epsilon^{ABC}A_{BCi}$ we can fix $A^A_i=0$
by a gauge tranformation. To do this fix a flat background
metric $\delta_{AB}$ and choose a 
$\lambda_{Ci}=\delta_{CD}\epsilon^{DEF}\partial_E \rho_{Fi}$
for some $\rho_{Ai}$.  Then under a gauge transformation we have
\f
A^{A \prime}_i = 0= A^A_i + \partial_i\epsilon^{ABC} \lambda_{BC}
+ \partial^A\partial^E \rho_{Ei} - \nabla^2 \rho_{Ai}
\ff
This can be solved locally $\rho_{Ai}$ of the form
\f
\rho_{Ai} = {1 \over \nabla^2} \left (
A^A_i + \partial_i\epsilon^{ABC} \lambda_{BC}
+ \partial^A\partial^E \rho_{Ei}
\right )
\ff
Similarly, we can fix the mixed components 
$A_{i\bar{A}\bar{B}}=0$ by using the gauge
transoformations parameterized by $\lambda_{\bar{C}i}$.  

To proceed further we reduce the degrees of freedom by
making an ansatz.  We dimensionally reduce by 
imposing that 
\f
\partial_A A_{abc}=\partial_{\bar{A}}A_{abc}=0
\label{ansatz1}
\ff
so that there is only spatial dependence on the $z$.  Then we
set the remaining mixed components to zero by imposing
\f
A_{iA\bar{A}}=A_{ijA}=A_{ij\bar{A}}=A_{AB\bar{A}}=
A_{A\bar{A}\bar{B}}=0
\label{ansatz2}
\ff

The remaining degrees of freedom are the components
${\cal A}={1\over 6} A_{ABC}\epsilon^{ABC}$
and $\bar{\cal A}={1 \over 6}
A_{\bar{A}\bar{B}\bar{C}}\epsilon^{\bar{A}\bar{B}\bar{C} }$. 
The non-vanishing components of $F_{abcd}$ are
$F_{ijkl}$ and $F_{iABC}= \partial_i {\cal A} \epsilon_{ABC}$
and $F_{i\bar{A}\bar{B}\bar{C}}= \partial_i 
\bar{\cal A} \epsilon_{\bar{A}\bar{B}\bar{C}}$.

All of the constraints (\ref{H}) vanish automatically under
this dimensional reduction except,
\f
H^{ij}=\epsilon^{ijkl}\partial_k {\cal A} \partial_l \bar{\cal A}=0 .
\label{newh}
\ff

The  symplectic form, in the presence of the dimensional
reduction is then,
\f
\omega (\delta_1 A, \delta_2 A )= \int \left (  
 \delta_1 {\cal A} \wedge \delta_2 \bar{\cal A}\tilde{\Phi}
- (1 \rightarrow 2)
\right )
\label{symfinal}
\ff
We can now invert to find the Poisson brackets,
\f
\{ {\cal A}(y,z) , \bar{\cal A}(y^\prime ,z^\prime) \}
= {1 \over  \tilde{\Phi}} \delta^6 (y,y^\prime ) \delta^4 (z,z^\prime )
\ff
 
Once we have these we can write the algebra for the gauge
invariant observables.  The fiber observables algebra works
as before, except that now the $\phi_I(z)$ 
and $\tilde{\pi}^J$ do depend on
$z$.  But we have to remember also the remaining constraint
(\ref{newh}).  One set of solutions follows if we set
$\partial_i{\cal A}=0$.  Then the remaining degrees of freedom
are the
\f
\phi_I(z)=e^{\int_{\gamma^I}\bar{\cal A}
\epsilon_{\bar{A}\bar{B}\bar{C}}}
\ff
which become functions on ${\cal Z}$.

By choosing representatives
of  the
$\pi^I$  that only involve the ${\cal A}$ we then
find the 
observables algebra,
\f
\{  \phi_I (z) , {\Pi}^J  \} = \delta_I^J  \phi_I (z)
\label{algebra3}
\ff
 
Thus, by dimensional reduction we have arrived at a 
reduction of the theory that has 
a gauge invariant observables algebra.  
We see that
the resulting structure is a new kind of combination of a
conventional and topological quantum field theory, which we
may call quasi-topological.
We have an infinite set of coordinate observables, who are
defined as local fields on a lower dimensional submanifold.
Each of them is connected with homotopy classes of the fiber
spaces.  In this sense we have something like a field theory
of topological field theories\footnote{This 
kind of structure was found for the five
dimensional abelian Chern-Simons theory by \cite{focketal}.}. 
The structure of the conjugate momenta are 
even more unusual, as we have only a finite set of 
distinct momenta who are associated to homotopy classes of
the fibers.  

It is clear that this kind of structure has arisen because of
we have considered a truncation of the theory defined by 
the dimensional reduction given by (\ref{ansatz1})
and (\ref{ansatz2}).  The full
set of solutions is likely to be even more intricate.  Steps
towards the construction of the general canonical formalism
are given in \cite{maxmark}.  For the present we confine
ourselves to a discussion of the quantum theory associated with
this quasi-topological sector.

\section{Quantum theory of the $4+6$ sector}

We may now proceed to a sketch of the quantum theory associated
to the sector of the theory we have just defined.  
In fact, since the topology of $\cal M$ is fixed before the
quantization we have one theory for each 
compact six manifold $\cal Y$.  In the last section we
concluded that if the homology has
a basis of  $N$ pairs 
$(\gamma_I, \pi^I)$, we have an observables algebra 
(\ref{algebra}) consisting of $N$ pairs of fields and
momenta on the four manifold $\cal Z$, given by (\ref{algebra3}). 
 
We are interested in a representation
of (\ref{algebra3}) on which we can also have a representation
of $Diff({\cal Z})$, so we can mod out by the diffeomorphisms
to find the physical states.  We can thus not use a standard Fock
representation.  As in the construction of the loop representation
we have to first construct a non-separable state space on
which $Diff({\cal Z})$ has an unbroken action.  To do this we
consider the $\phi_I (z)$ to be creation operators that create
an excitation, which corresponds to a surface in the $I$'th
homology class in the fiber $\cal Y$ over the point 
$z \in {\cal Z}$. That is we take the continuous product of
Fock spaces over each point $z$.  Thus, we have a vacuum
state $|0>$ defined by $\tilde{\pi}(z) |0>=0$.  We then define
states by occupation numbers at each point $z$.  Thus, a
general state consists of a finite list of of excitations
$|(I_1,z_1)....(I_n ,z_n )>$ where each pair $(I,z)$ represents
a surface in the $I$'th homology  created at the point $z$.
The action of the field operators is then to add excitations
\f
\hat{\phi}_I (z) |(I_i,z_i )> = | (I,z) \cup (I_i , z_i )>
\ff
The Hilbert space is then simply
\f
{\cal H}_{kin}= \prod_z {\cal H}_z
\label{kinematical}
\ff
where a basis for the  Hilbert space at each point
${\cal H}_z$ is labeled by $|n_1,...,n_N >$, where
the $n_i$'s are the occupation numbers
in each of the homology classes 
$\tilde{\gamma}_I$.  A state is then given by
\f
|\Psi>= \otimes_z |\psi , z>
\ff
where $|\psi , z> \in {\cal H}_z$.  
The inner product is then simply the product,
\f
<\Psi|\Psi^\prime> = \prod_z <\psi ,z |\psi^\prime , z>
\label{ip}
\ff
The normalizable states are those in which there are
only a finite  number of excitations, so that only a
finite number of the factors in (\ref{ip}) differ from
one.

The action of the conjugate momentum variables is defined by,
\f
\hat{\Pi}^J |(I_1,z_1)....(I_n ,z_n )>=
\delta^J_{I_i}
|(I_1,z_1)..(I_{i-1},z_{i-1}) (I_{i+1},z_{i+1})..(I_n ,z_n )>
\ff
i.e. the operator acts to remove the excitations in $\tilde{\gamma}_J$.

The space (\ref{kinematical}) is non-separable.  But it is easy
to mod out by diffeomorphisms.  
To do this we may define a unitary representation of 
$Diff({\cal Z})$ on ${\cal H}_{kin}$.  Given $\phi \in Diff({\cal Z})$
we may define
\f
U(\phi ) |(I_i,z_i )>= |(I_i,\phi^{-1} \cdot z_i )>
\ff
It is straightforward to check that this operator is unitary.  As in the 
case of the loop representation, there are no anomalies of
the diffeomorphism group.  
Diffeomorphism invariant states are then defined by 
\f
\Psi [(I_i,z_i )] = <\Psi | (I_i,z_i )> = \Psi [(I_i,\phi^{-1} \cdot z_i )]
\ff
The resulting diffeomorphism
invariant Hilbert space ${\cal H}^{diffeo}$ is labeled simply
by a basis of states corresponding to excitations at 
distinct points, with no labelings as to where the points are
in $\cal Z$.  Thus, a basis is given by a finite set of $P$  lists
\f
|(n_i)_1 , ..., (n_i)_P>
\ff
where each list $n_i= (n_1, ...,n_N)$ consists of occupation numbers
for the $N$ homology classes $\tilde{\gamma}_I$.  Thus, 
${\cal H}^{diffeo}$ has a separable basis. A non-separable Hilbert
space was just needed as a technical device at the kinematical
level, as in non-perturbative quantum gravity.
Diffeomorphism invariant observables exist, such as the number
operators
\f
N^I \equiv \int_{\cal Z} \phi_I (z) p^I
\ff
(where no sum on $I$ is taken.)  More complicated operators
may be easily constructed, which measure how many points
there are at which there is a certain pattern of excitations.

As there are no interesting diffeomorphism classes of sets of
points, the diffeomorphism invariant quantum theory is in this
case rather boring.  We see that there are local degrees of
freedom in the four dimensional manifold $\cal Z$
that do correspond to three dimensional surfaces wrapped 
around various homology classes of the six manifold
$\cal Y$.  But there are no interesting relationships or
interactions among them. 
The structure of the other cases in which
$d>4$ are more interesting, as there are non-trivial extended
structures in both $\cal Z$ and $\cal Y$.  These will be
discussed elsewhere.

\section{Conclusions}

To summarize, we have found a sector of the solution space
of the theory which corresponds to bundles defined by fibering
six dimensional compact manifolds $\cal Y$ over a four manifold
$\cal Z$.  The degrees of freedom  are a canonically conjugate
pair of fields $(\phi_I (z), \tilde{\Pi}^J)$ on 
$\cal Z$ corresponding to a basis of the third
homology of $\cal Y$. These have a quasi-topological structure
in that the $\phi_I (z)$ are local fields on
$\cal Z$ while the momenta $\Pi^I$
depend only on the homology classes. The observables algebra
 is given entirely in terms
of the intersection numbers of the surfaces in $\cal Y$.

Thus, we have achieved our goal of finding a sector of the
theory corresponding to a natural compactification of the theory
in which the geometry of the compactified directions is entirely
represented by an algebra of functions on the base manifold.

Furthermore, we have constructed the quantum theory 
associated to this sector of the theory, and discovered it
consists of diffeomorphism classes (in the four dimensional
manifold $\cal Z$) of excitations of wrappings of three surfaces
around homology classes in the six dimensional manifold
$\cal Y$.

\section*{ACKNOWLEDGEMENTS}

I am especially grateful to Eli Hawkins for seeing an error
in the original version of this paper. In addition, 
I would like to thank John Baez, Louis Crane,  
Fotini Markopoulou and 
Andrew Strominger for discussions and encouragement.
Comments on the manuscript by Shyamoli Chaudhuri and 
Juan Maldecena were also very helpful.This work was 
supported by NSF grant PHY-9514240 to The Pennsylvania State
University and a NASA grant to The Santa Fe Institute.

\end{document}